\def\year{2019}\relax
\documentclass[letterpaper]{article} 
\usepackage{aaai19}  
\usepackage{times}  
\usepackage{helvet}  
\usepackage{courier}  
\usepackage{url}  
\usepackage{graphicx}  
\newcommand{\namecite}[1]{\citeauthor{#1}~(\citeyear{#1})}

\begin{document}
%
\title{Personalized Education at Scale}
\author{Sam Saarinen, Evan Cater, and Michael Littman\\
Brown University\\
Providence, Rhode Island, USA\\
Contact: sam-saarinen.github.io\\
}

\maketitle

\section{Personalized Education}

Tailoring the presentation of information to the needs of individual students leads to massive gains in student outcomes~\cite{bloom19842}. This finding is likely due to the fact that different students learn differently, perhaps as a result of variation in ability, interest or other factors~\cite{schiefele1992interest}.
Adapting presentations to the educational needs of an individual has traditionally been the domain of experts, making it expensive and logistically challenging to do at scale, and also leading to inequity in educational outcomes.

Increased course sizes and large MOOC enrollments provide an unprecedented access to student data.  We propose that emerging technologies in reinforcement learning (RL), as well as semi-supervised learning, natural language processing, and computer vision are critical to leveraging this data to provide personalized education at scale.

\section{Sources of Difficulty}

Personalized instruction is readily cast as a reinforcement-learning problem.  The student's knowledge and interests are a (sometimes unobserved) state, the collection of pedagogical tools (videos, text, activities, games, etc.) are the set of actions with corresponding costs reflecting their demands in terms of time or other resources, and performance on some measure of learning (say a final exam) is the final reward.\footnote{There has been some concern in contemporary educational circles about ``Teaching to the Test".  This problem is familiar to RL researchers who have observed that handcrafted rewards can lead to unintended behavior~\cite{dario2016faulty}. On the whole, education research has been moving towards better-validated assessments, which have gained traction as standardized tools such as the Force Concept Inventory~\cite{hestenes1992force}, ACS-Standardized General Chemistry exam, or AP exams.}

The largest barrier to automation is the scale of the problem, which comes into play in three ways.  First, a comprehensive student model may contain many features that are extraneous for any particular topic, resulting in increased sample complexity demands for learning.  Second, any given topic may have thousands of pedagogical actions associated with it.  (A video search for ``derivatives tutorial" yields about $10^5$ results.)  
Third, student state changes are typically unobserved, and there may be a large temporal delay between when an action is taken and when reward (learning demonstrated on a test) is observed.  Human instructors frequently use ad-hoc informal tests and affective impressions to gauge the effect of educational content, but replicating this observational adaptability is challenging.

\section{Current Progress}

Intelligent tutoring systems have been successfully deployed in a number of problem-rich domains, such as LISP programming, Algebra, and Genetics~\cite{koedinger2006cognitive}.  These systems achieve near human-tutor level gains~\cite{corbett2001cognitive} by inferring student mastery (state) over practice problems, then using an expert-crafted policy to advance students through a linear curriculum.  
These techniques are currently difficult and expensive to apply to less structured domains, and they do not accommodate much variation in students.

It has been previously proposed that RL could be used to optimize pedagogical approaches~\cite{iglesias2009learning}.  Unfortunately, the RL algorithms previously considered do not scale well to teaching an entire course.  For example, \namecite{rafferty2011faster} explore the possibility of using a POMDP to model student belief over several possible conceptions of a single task.  Their empirical study shows the efficacy of adaptation to inferred state, but the technique is not currently amenable to larger-scale problems. \namecite{chi2011empirically} consider introductory physics tutoring as an RL problem, but are limited to modeling only a few actions and state features because of the cost of collecting exploratory trajectories under a uniform random policy.  Similar work in computer science education~\cite{iglesias2009learning} used simulated students to collect data for an initial policy, but were also highly constrained because of the high sample complexity of Q learning.  In the next three sections, we propose problem formulations and the application of emerging techniques that could overcome these barriers to scalability.



\section{A Contextual Bandit Problem}
Assume that for each student we have explicitly constructed a feature vector (say by administering personality tests, clustering based on behavior in previous courses, or through some kind of tagging system), and we have a collection of possible pedagogic actions.  If we look at only immediate  performance on a topical assessment, we can cast the personalized instruction problem as a contextual bandit.  Approaches such as the contextual Gaussian process bandit algorithm~\cite{srinivas2012information} suggest that this kind of personalization may be feasible, although this approach does not solve the longer-term problem of curricular planning.  To achieve sufficiently low sample complexity, an efficient task-relevant characterization of contexts and actions is necessary, requiring a compact model of students and a similarity measure between pedagogical resources.

\section{A (Hierarchical) POMDP}

In many educational settings, it is impractical to assess students after every pedagogical action due to the expense in creation of validated assessments, administration, and student time.  As a result, 
reward is delayed and state is only partially observable.  Additionally, we do not have access to the ``true" model of the student.  In this setting, relevant hidden features must be invented and inferred by the algorithm.  The scale of the curricular problem may be mitigated by decomposing the curriculum into shorter term goals, akin to an instructor's ``units".  There has been some compelling work in the direction of decomposing POMDPs: the work of \namecite{wray2017online} decomposes a large problem into entity-specific POMDPs whose recommendations are combined, and \namecite{sridharan2010planning} decomposes observational actions at different levels of granularity.  Approaches like these may allow efficient (approximate) solution of otherwise intractably large POMDPs.

\section{An ``Active" POMDP with Human-Collaborative Actions}

Eventually, we desire a system that can not only personalize instruction autonomously, but can also collaborate with human experts.  In particular, we would like such a system to identify pedagogical bottlenecks where new pedagogical actions may have greater efficacy for a specific subset of students, or to identify states of high uncertainty where a targeted assessment might differentiate hidden states.  This idea is similar in spirit to the work of \namecite{mandel2017add}, which considers the problem of finding the optimal states for which to ask a human (expert) to construct new actions.

\section{Integration with Other Techniques}

While the underlying problems are defined in reinforcement-learning terms, their solutions will likely integrate some of the cutting-edge techniques from other areas.  For example, semantic embeddings developed in the context of Natural Language Processing may provide a way to generalize educational insights across enormous action spaces.  Computer vision and emotion detection could play a valuable role in reading students and reducing uncertainty of their state. Techniques from semi-supervised learning might be used to improve the generalization across student states.  Recent work in explainable machine learning may improve adoption and use of personalized instruction agents, and also provide insight into effective student--pedagogy combinations.

\section{Call to Action}

We have an unprecedented access to student data at scale, exciting new developments in scalable RL, and compelling deployed technologies in natural language processing and computer vision.  If the AI research community invests time into researching sample-efficient RL algorithms as outlined above, the impact to education---as well as many other domains that would benefit from interactive personalization---would be profound and far reaching.

\bibliographystyle{aaai}
\bibliography{RLed}

\begin{thebibliography}{}

\bibitem[\protect\citeauthoryear{Bloom}{1984}]{bloom19842}
Bloom, B.~S.
\newblock 1984.
\newblock The 2 sigma problem: The search for methods of group instruction as
  effective as one-to-one tutoring.
\newblock {\em Educational researcher} 13(6):4--16.

\bibitem[\protect\citeauthoryear{Chi \bgroup et al\mbox.\egroup
  }{2011}]{chi2011empirically}
Chi, M.; VanLehn, K.; Litman, D.; and Jordan, P.
\newblock 2011.
\newblock Empirically evaluating the application of reinforcement learning to
  the induction of effective and adaptive pedagogical strategies.
\newblock {\em User Modeling and User-Adapted Interaction} 21(1-2):137--180.

\bibitem[\protect\citeauthoryear{Corbett}{2001}]{corbett2001cognitive}
Corbett, A.
\newblock 2001.
\newblock Cognitive computer tutors: Solving the two-sigma problem.
\newblock In {\em International Conference on User Modeling},  137--147.
\newblock Springer.

\bibitem[\protect\citeauthoryear{Dario}{2016}]{dario2016faulty}
Dario, J.
\newblock 2016.
\newblock Faulty reward functions in the wild.
\newblock \url{https://blog.openai.com/faulty-reward-functions/}.

\bibitem[\protect\citeauthoryear{Hestenes, Wells, and
  Swackhamer}{1992}]{hestenes1992force}
Hestenes, D.; Wells, M.; and Swackhamer, G.
\newblock 1992.
\newblock Force concept inventory.
\newblock {\em The physics teacher} 30(3):141--158.

\bibitem[\protect\citeauthoryear{Iglesias \bgroup et al\mbox.\egroup
  }{2009}]{iglesias2009learning}
Iglesias, A.; Mart{\'\i}nez, P.; Aler, R.; and Fern{\'a}ndez, F.
\newblock 2009.
\newblock Learning teaching strategies in an adaptive and intelligent
  educational system through reinforcement learning.
\newblock {\em Applied Intelligence} 31(1):89--106.

\bibitem[\protect\citeauthoryear{Koedinger and
  Corbett}{2006}]{koedinger2006cognitive}
Koedinger, K.~R., and Corbett, A.~T.
\newblock 2006.
\newblock Cognitive tutors: technology bringing learning science to the
  classroom.

\bibitem[\protect\citeauthoryear{Mandel \bgroup et al\mbox.\egroup
  }{2017}]{mandel2017add}
Mandel, T.; Liu, Y.-E.; Brunskill, E.; and Popovic, Z.
\newblock 2017.
\newblock Where to add actions in human-in-the-loop reinforcement learning.
\newblock In {\em AAAI},  2322--2328.

\bibitem[\protect\citeauthoryear{Rafferty \bgroup et al\mbox.\egroup
  }{2011}]{rafferty2011faster}
Rafferty, A.~N.; Brunskill, E.; Griffiths, T.~L.; and Shafto, P.
\newblock 2011.
\newblock Faster teaching by pomdp planning.
\newblock In {\em International Conference on Artificial Intelligence in
  Education},  280--287.
\newblock Springer.

\bibitem[\protect\citeauthoryear{Schiefele, Krapp, and
  Winteler}{1992}]{schiefele1992interest}
Schiefele, U.; Krapp, A.; and Winteler, A.
\newblock 1992.
\newblock Interest as a predictor of academic achievement: A meta-analysis of
  research.

\bibitem[\protect\citeauthoryear{Sridharan, Wyatt, and
  Dearden}{2010}]{sridharan2010planning}
Sridharan, M.; Wyatt, J.; and Dearden, R.
\newblock 2010.
\newblock Planning to see: A hierarchical approach to planning visual actions
  on a robot using pomdps.
\newblock {\em Artificial Intelligence} 174(11):704--725.

\bibitem[\protect\citeauthoryear{Srinivas \bgroup et al\mbox.\egroup
  }{2012}]{srinivas2012information}
Srinivas, N.; Krause, A.; Kakade, S.~M.; and Seeger, M.~W.
\newblock 2012.
\newblock Information-theoretic regret bounds for gaussian process optimization
  in the bandit setting.
\newblock {\em IEEE Transactions on Information Theory} 58(5):3250--3265.

\bibitem[\protect\citeauthoryear{Wray, Witwicki, and
  Zilberstein}{2017}]{wray2017online}
Wray, K.~H.; Witwicki, S.~J.; and Zilberstein, S.
\newblock 2017.
\newblock Online decision-making for scalable autonomous systems.
\newblock In {\em Proceedings of the 26th International Joint Conference of
  Artificial Intelligence (IJCAI)},  4768--4774.

\end{thebibliography}

\end{document}